\documentclass[twocolumn,showpacs,prl]{revtex4}

\usepackage{epsf}
\usepackage{amssymb}

\begin{document}

\title{Resonance- and chaos-assisted tunneling in mixed regular-chaotic 
  systems}

\author{Christopher Eltschka}
\author{Peter Schlagheck}

\affiliation{Institut f{\"u}r Theoretische Physik, Universit{\"a}t Regensburg,
  93040 Regensburg, Germany}

\date{\today}

\begin{abstract}

We present evidence that nonlinear resonances govern the tunneling process
between symmetry-related islands of regular motion in mixed regular-chaotic
systems.
In a similar way as for near-integrable tunneling, such resonances induce
couplings between regular states within the islands and states that are
supported by the chaotic sea.
On the basis of this mechanism, we derive a semiclassical expression for the
average tunneling rate, which yields good agreement in comparison with the
exact quantum tunneling rates calculated for the kicked rotor and the kicked
Harper.

\end{abstract}


\maketitle

Despite its genuine quantal character, dynamical tunneling \cite{DavHel81JCP}
is strongly sensitive to details of the underlying classical phase space
\cite{Cre98}.
A particularly prominent scenario in this context is ``chaos-assisted''
tunneling \cite{LinBal90PRL,BohTomUll93PR,TomUll94PRE,DorFri95PRL}
which takes place between quantum states that are localized on two
symmetry-related regular islands in a mixed regular-chaotic phase space.
The presence of an appreciable chaotic layer between the islands dramatically
enhances the associated tunneling rate as compared to the integrable case, and
induces strong fluctuations of the rate at variations of external parameters
\cite{LinBal90PRL,BohTomUll93PR}.
This phenomenon is attributed to the influence of ``chaotic states'' that are
distributed over the stochastic sea.
Since such chaotic states typically exhibit an appreciable overlap with the
boundary regions of both islands, they may provide efficient ``shortcuts''
between the two regular quasimodes in the islands
\cite{BohTomUll93PR,TomUll94PRE,DorFri95PRL}.
Indeed, chaos-assisted tunneling processes arise in a number of physical
systems, e.g.\ in the ionization of resonantly driven hydrogen 
\cite{ZakDelBuc98PRE}, in microwave or optical cavities
\cite{NoeSto97N,DemO00PRL}, as well as in the effective pendulum dynamics
describing tunneling experiments of cold atoms in optical lattices
\cite{MouO01PRE,HenO01N}.

While the statistical properties of the chaos-assisted tunneling rates are
well reproduced by a random matrix description of the chaotic part of the
Hamiltonian \cite{LeyUll96JPA}, the formulation of a tractable and reliable
semiclassical theory for the average tunneling rate is still an open problem.
Promising progress in this direction was reported by Shudo and coworkers 
\cite{ShuIke95PRL} who obtain a good quantitative reproduction of classically
forbidden propagation processes in mixed systems by incorporating complex
trajectories into the semiclassical propagator 
Their approach requires, however, the study of highly nontrivial structures in
complex phase space, and cannot be straightforwardly connected to single
coupling matrix elements between regular and chaotic states.
A complementary ansatz, based on a Bardeen-type expression for the coupling
to the chaos, was presented by Podolsky and Narimanov \cite{PodNar03PRL}.
In comparison with tunneling rates from the driven pendulum, good agreement
was obtained for large and moderate $\hbar$, whereas significant deviations seem
to occur deep in the semiclassical regime \cite{PodNar03PRL}.

In the present Letter, we shall point out that {\em nonlinear resonances}
between different classical degrees of freedom play a crucial role in
chaos-assisted tunneling processes.
Such nonlinear resonances are known to govern tunneling between
symmetry-related wells in near-integrable systems
\cite{Ozo84JPC,BonO98PRE,BroSchUll01PRL,Kes03JCP}, where they
induce transitions to highly excited states inside the well and thereby
strongly enhance the tunneling rate \cite{BroSchUll01PRL}.
We shall argue that the same mechanism is also responsible for the
semiclassical coupling between regular and chaotic states in mixed systems,
and determines the average tunneling rate in chaos-assisted tunneling.
A simple semiclassical expression derived from this principle shows indeed
reasonably good agreement with the exact quantum splittings.

We restrict our study to systems with one degree of freedom that evolve under
a periodically time-dependent Hamiltonian $H(p,q,t) = H(p,q,t + \tau)$ and are
visualized by a stroboscopic Poincar{\'e} section evaluated at $t = n \tau$ 
($n \in \mathbb{Z}$).
We suppose that $H$ possesses a discrete symmetry which, for a suitable choice
of a parameter of $H$, leads to a mixed phase space with two symmetric regular
islands that are separated by a chaotic sea.
We furthermore assume that each of the symmetric islands exhibits a prominent
$r$:$s$ resonance---i.e., where $s$ internal oscillations around the
island's center take place within $r$ periods of the driving---which manifests
itself in the stroboscopic section as a chain of $r$ sub-islands that are
embedded in the torus structure of the regular island.

The motion in the vicinity of the $r$:$s$ resonance is approximately
integrated by secular perturbation theory \cite{LicLie}.
For this purpose, we formally introduce a time-independent Hamiltonian
$H_0(p,q)$ that approximately reproduces the regular motion in the islands,
and denote by $(I,\theta)$ the action-angle variables describing the dynamics
within each of the islands.
After the canonical transformation $\theta \mapsto \vartheta = \theta - s/r \cdot 2\pi t/\tau$ to the
frame that corotates with the resonance, and after averaging the
resulting Hamiltonian over $r$ periods of the external driving, which
is justified since $\vartheta$ varies slowly with time near resonance, we
obtain in lowest nonvanishing order 
\begin{equation}
  H_{\rm eff}(I,\vartheta) = \frac{(I - I_{r:s})^2}{2 m_{r:s}} + 2 V_{r:s} \cos r \vartheta
  \label{eq:heff}
\end{equation}
as effective integrable Hamiltonian for the dynamics near the resonance.
Here, $I_{r:s}$ denotes the action variable at resonance, 
$1/m_{r:s}$ parametrizes the variation of the internal oscillation frequency
with $I$ at resonance, and $V_{r:s}$ characterizes the strength of the
perturbation.

Comparing the pendulum-like dynamics of this effective Hamiltonian with the
actual classical dynamics generated by $H$ provides an access to the
parameters of $H_{\rm eff}$ without explicitly using the functional form of
$H_0(p,q)$.
To this end, we numerically calculate the monodromy matrix $M_{r:s}$ of a
stable periodic point of the resonance (which involves $r$ iterations of the
stroboscopic map) as well as the phase space areas $S_{r:s}^+$ and $S_{r:s}^-$
that are enclosed by the outer and inner separatrices of the resonance,
respectively.
Using the fact that the trace of $M_{r:s}$ as well as the phase space areas
$S_{r:s}^\pm$ remain invariant under the canonical transformation to $(I,\vartheta)$, we
infer
\begin{eqnarray}
  I_{r:s} & = & \frac{1}{4 \pi} ( S_{r:s}^+ + S_{r:s}^- ) \, , \label{eq:area} \\
  \sqrt{2 m_{r:s} V_{r:s}} & = & \frac{1}{16} ( S_{r:s}^+ - S_{r:s}^- ) \, ,
  \label{eq:sep} \\
  \sqrt{\frac{2 V_{r:s}}{m_{r:s}}} & = & \frac{1}{r^2 \tau} \arccos({\rm tr} \,
  M_{r:s}/2) \label{eq:trm}
\end{eqnarray}
from the integration of the dynamics generated by $H_{\rm eff}$, which allows
us to determine $I_{r:s}$, $m_{r:s}$, and $V_{r:s}$.

The implications of the nonlinear resonance for the corresponding quantum
system can be directly seen from the representation of the quantized version
of $H_{\rm eff}$ in the eigenbasis of $H_0$, which consists of ``even'' and
``odd'' functions with respect to the discrete symmetry of $H$.
In the action-angle variable representation, the eigenfunctions of $H_0$ are,
for a fixed parity, essentially given by plane waves $\psi_n(\vartheta) \sim \exp(i n \vartheta)$ as
a function of the angle variable, where the integer index $n$ denotes the
excitation as counted from the center of the island.
The first, ``kinetic'' term of $H_{\rm eff}$ is therefore diagonal in this
basis with the matrix elements 
\begin{equation}
  E_n = [\hbar(n + 1/2) - I_{r:s}]^2/(2m_{r:s}) \, , \label{eq:en}
\end{equation}
while the ``potential'' term $2 V_{r:s} \cos r \vartheta$ induces couplings between
$\psi_n$ and $\psi_{n\pm r}$ with the matrix element $V_{r:s}$.
In this way, a perturbative chain is created that connects the ``ground
state'' $\psi_0$ of the island to the excited states $\psi_{lr}$ with integer $l$.
As was shown in Ref.~\cite{BroSchUll01PRL}, this coupling
mechanism generally leads to a strong enhancement of the level splitting
between the even and the odd ground state in the near-integrable regime, since
the unperturbed tunneling rate of a highly excited state $\psi_{lr}$ is much
larger than that of $\psi_0$.

In the mixed regular-chaotic case, the above tridiagonal structure of the
effective Hamiltonian becomes invalid beyond a maximum excitation index $n_c$
that marks the chaos border, i.e.\ for which $2\pi \hbar (n_c + 1/2)$ roughly
equals the size of the island.
Basis states $\psi_n$ with $n > n_c$ are defined on tori of $H_0$ that are
destroyed by the presence of other strong resonances, and therefore exhibit on
average a more or less equally strong coupling to each other. 
In the simplest possible approximation, which neglects the presence of partial
barriers in the chaos \cite{BohTomUll93PR}, the ``chaotic block''
$(H_{n,n'})_{n,n' > n_c}$ of the effective Hamiltonian is therefore
represented by a random matrix from the Gaussian orthogonal ensemble
\cite{TomUll94PRE,LeyUll96JPA}.

The probability density $P(\Delta E)$ for obtaining the level splitting $\Delta E$
between the ground state energies of the two symmetry classes can now be
calculated by performing the random matrix average over the chaotic part of
the Hamiltonian.
As was worked out by Leyvraz and Ullmo \cite{LeyUll96JPA}, this leads to a
Cauchy distribution
\begin{equation}
  P(\Delta E) = \frac{4 N_c \Delta_c V_{\rm eff}^2}{(N_c \Delta_c \, \Delta E)^2 + 4 \pi^2 V_{\rm eff}^4} 
  \label{eq:peff}
\end{equation}
with a cutoff at $\Delta E \sim 2 V_{\rm eff}$, where $N_c$ and $\Delta_c$ denote the
number of chaotic states and their mean level spacing at energy $E_0$,
respectively, and $V_{\rm eff}$ represents the effective coupling matrix
element between the ground state and the chaotic block.
In the presence of the nonlinear resonance inside the island, the latter is
evaluated by means of the tridiagonal structure within the regular part of the
Hamiltonian:
assuming $V_{r:s}$ to be much smaller than the intermediate energy
differences, we obtain
\begin{equation}
  V_{\rm eff} = V_{r:s} \prod_{l=1}^{k - 1}\frac{V_{r:s}}{E_0 - E_{lr}} \label{eq:veff}
\end{equation}
where the energies $E_{lr}$ are computed from Eq.~(\ref{eq:en}).
The elimination of intermediate regular states is performed up to the first
state $\psi_{kr}$ that is already located beyond the chaos border [i.e.,
$(k-1)r < n_c < kr$].

Since tunneling rates and their parametric variations are typically studied in
a logarithmic representation, the relevant quantity to be calculated from
Eq.~(\ref{eq:peff}) and compared to quantum data is not the mean value of $\Delta E$
(which would diverge if the cutoff is not taken into account), but rather the
average of the {\em logarithm} of $\Delta E$.
We therefore obtain the ``mean'' level splitting $\overline{\Delta E}$ as
\begin{equation}
  \overline{\Delta E} \equiv \frac{V_{\rm eff}^2}{N_c \Delta_c} \exp \left( \left\langle \log
      \frac{N_c \Delta_c \, \Delta E}{V_{\rm eff}^2} \right\rangle \right) 
  = \frac{2 \pi V_{\rm eff}^2}{N_c \Delta_c}
\end{equation}
where $\langle\ldots\rangle$ denotes the average with respect to the probability distribution
(\ref{eq:peff}).
The expression for the mean splitting further simplifies for our case of
periodically driven systems, where the eigenphases of the time evolution
operator are calculated.
Using the fact that the chaotic eigenphases are more or less uniformly
distributed in the interval $0 \leq \varphi < 2 \pi$, we obtain $N_c \Delta_c = \hbar \omega = 2 \pi \hbar / \tau$.
This results in the mean eigenphase splitting
\begin{equation}
  \overline{\Delta \varphi} = \frac{\tau \overline{\Delta E}}{\hbar} = 
  \left( \frac{\tau V_{\rm eff}}{\hbar} \right)^2 \, .
\end{equation}

\begin{figure}[t]
\epsfxsize8cm
\epsfbox{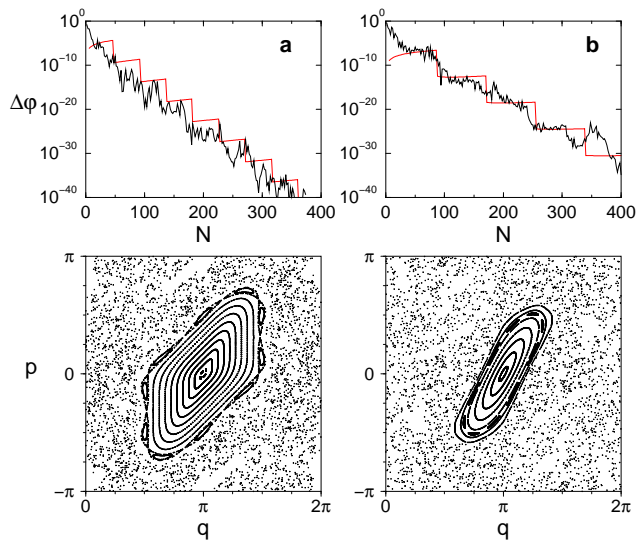}
\caption{
  Chaos-assisted tunneling in the kicked rotor at (a) $K = 2$ and (b) $K = 3$.
  Plotted are, as a function of $N = 2 \pi / \hbar$, the splittings $\Delta \varphi$ between
  the eigenphases of the ground states at the Bloch phases $\xi = 0$ and $\pi$.
  The associated regular islands are shown in the lower panels.
  The step-like curve is the semiclassical prediction $\overline{\Delta \varphi}$ for the
  eigenphase splittings, taking into account the 10:2 resonance at $K = 2$ and
  the 10:3 resonance at $K = 3$.
  \label{fg:spkr}
}
\end{figure}

\begin{figure}[t]
\epsfxsize8cm
\epsfbox{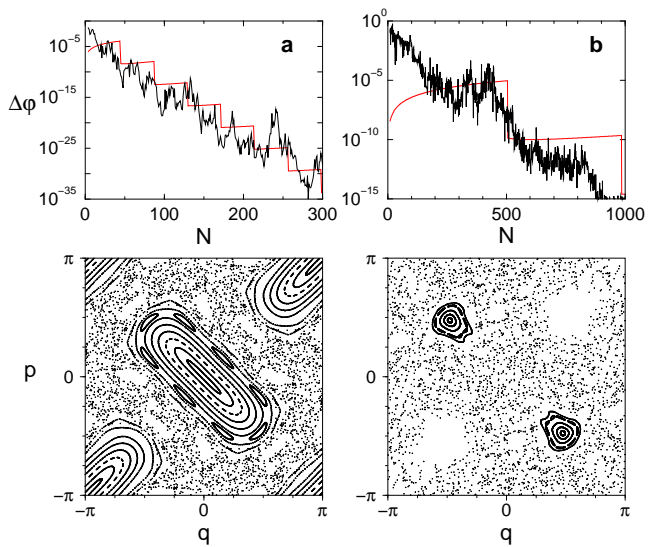}
\caption{
  Chaos-assisted tunneling in the kicked Harper at (a) $\tau = 2$ and (b)
  $\tau = 3$.
  Plotted are in (a) the splittings between the eigenphases of the ground
  state at $\xi = 0$ and $\pi$, and in (b) the eigenphase splittings between the
  symmetric and the antisymmetric states that are localized on the center of
  the small, bifurcated islands (at fixed $\xi = 0$).
  The step-like curves represent the semiclassical predictions, based on the
  8:2 resonance at $\tau = 2$ and on the 9:1 resonance at $\tau = 3$.
  \label{fg:spkh}
}
\end{figure}

To illustrate our theory, we apply it to one-dimensional systems that are
subject to time-periodic kicks.
Their classical Hamiltonian is given by
\begin{equation}
  H(p,q,t) = T(p) + \sum_{n=-\infty}^\infty \tau \delta(t - n \tau) V(q)
\end{equation}
where $T(p)$ and $V(q)$ denote the kinetic energy and the potential associated
with the kick, respectively.
The time evolution can be represented by the map $(p,q) \mapsto
(\tilde{p},\tilde{q})$ with $\tilde{p} = p - \tau V'(q)$ and 
$\tilde{q} = q + \tau T'(p)$, which describes the stroboscopic Poincar{\'e} section
at times immediately before the kick.
The corresponding quantum dynamics is generated by the unitary operator
\begin{equation}
  U = \exp\left( - \frac{i \tau}{\hbar} T(\hat{p}) \right)
  \exp\left( - \frac{i \tau}{\hbar} V(\hat{q}) \right)
\end{equation}
where $\hat{p}$ and $\hat{q}$ denote the momentum and position operator,
respectively.

Specifically, we consider the kicked rotor given by $T(p) = \frac{1}{2} p^2$
and $V(q) = K \cos q$ with $\tau \equiv 1$, and the kicked Harper given by 
$T(p) = \cos p$ and $V(q) = \cos q$ \cite{LebO90PRL}.
Using Bloch's theorem, we restrict our study to eigenfunctions of $U$ that are
periodic in position.
We furthermore choose $\hbar = 2 \pi / N$ where $N$ is an even integer.
This allows us, for both the kicked rotor and the kicked Harper, to write the
periodic eigenfunctions as Bloch functions in momentum---i.e., with
$\tilde{\psi}(p + 2 \pi) = \tilde{\psi}(p) \exp(i \xi)$ where $\tilde{\psi}$ is the Fourier
transform of $\psi$.
Since the subspace of such functions is $N$-dimensional for fixed $\xi$ 
\cite{LebO90PRL}, the eigenphases and eigenvectors of $U$ can be calculated by
diagonalizing finite $N \times N$ matrices.

Quantum tunneling can take place between a regular island in the fundamental
phase space cell and its periodically shifted counterparts.
As a consequence, different Bloch phases $\xi$ lead to slightly different
eigenphases for states that are localized on a given torus in the island.
The spectral quantity that we discuss in the following is the difference
$\Delta\varphi = |\varphi^{(\xi=0)} - \varphi^{(\xi=\pi)}|$ between the eigenphases of the island's ground
state for $\xi=0$ and $\xi=\pi$.
$\varphi^{(\xi=0)}$ and $\varphi^{(\xi=\pi)}$ are calculated by diagonalizing $U$ in a suitable
basis \cite{BroSchUll01PRL}, and by identifying the ground state from the
localization properties of the eigenstates near the center of the island. 
Multiple precision arithmetics is used in order to calculate splittings below
$\Delta\varphi \sim 10^{-15}$.

Figs.~\ref{fg:spkr}(a), \ref{fg:spkr}(b), and \ref{fg:spkh}(a) show the
eigenphase splittings for the kicked rotor and the kicked Harper,
respectively, as a function of $N = 2 \pi / \hbar$, calculated for $K = 2$
and $3$ in Fig.~\ref{fg:spkr} as well as for $\tau = 2$ in Fig.~\ref{fg:spkh}(a).
The step-like curves show our semiclassical predictions of the eigenphase
splittings, which are based on prominent resonance chains boldly marked in the
corresponding phase space.
The relevant parameters $m_{r:s}$, $V_{r:s}$ and $I_{r:s}$ are computed from
phase space areas and periodic points via Eqs.~(\ref{eq:area}--\ref{eq:trm}).
From the numerically calculated phase space area $S$ covered by the island, we
infer the number $k$ of intermediate steps that are necessary to couple the
ground state to the chaos.
An artificially sharp decrease of the semiclassical splitting $\overline{\Delta \varphi}$
therefore occurs whenever $\hbar$ passes through a value where 
$S = 2 \pi \hbar (kr + 0.5)$ with integer $k$.

In spite of the number of simplifications and approximations that are involved
in the derivation of the semiclassical expression for the mean eigenphase
splittings, we obtain a relatively good agreement between $\Delta\varphi$ and
$\overline{\Delta \varphi}$.
In particular, the first major plateau in the quantum splittings is remarkably
well matched by the semiclassical curve, which clearly indicates that the
coupling to the chaotic sea is mediated by the nonlinear resonance there.
Our method fails to reproduce the quantum splittings in the ``anticlassical''
limit of large $\hbar$, e.g.\ for $N < 50$ in Fig.~\ref{fg:spkr}(b).
Preliminary calculations show, however, that a better agreement in this regime
might be obtained by properly taking into account the action dependence of
$V_{r:s}$ in the effective Hamiltonian (\ref{eq:heff}), which is completely
neglected in the present treatment.
More details will be presented in a subsequent publication.

Apart from Fig.~\ref{fg:spkr}(b), where also plateaus of higher order are
well reproduced by the semiclassical theory, we observe a systematic tendency
to overestimate the exact quantum splittings for low and moderate values of
$1/\hbar$.
We tentatively attribute this fact to the existence of partial barriers in the
chaotic part of the phase space, which may enhance the effective size of the 
island for the quantum tunneling process.
In particular, it is known that ``Cantori'' (i.e., broken tori) in the chaos
inhibit the quantum transport in a similar way as invariant tori in the
island, as long as the phase space area associated with the classical flux
through the Cantorus is smaller than $\pi \hbar$ \cite{GeiRadRub86PRL}.
A better agreement with the quantum splittings might therefore be obtained by
properly incorporating {\em hierarchical} states \cite{KetO00PRL}, which are
localized in the immediate vicinity of the island, into the semiclassical
description (see in this context also \cite{DorFri95PRL}).

Finally, Fig.~\ref{fg:spkh}(b) shows the case of tunneling between two
symmetric regular islands in the kicked Harper at $\tau = 3$, which arise from a
bifurcation of the central island taking place at $\tau = 2$.
The quantum splittings are now given by the eigenphase difference between the
symmetric and the antisymmetric state associated with the pair of islands,
calculated here at fixed $\xi = 0$.
We see that the splittings display a prominent plateau at $N \simeq 300 \ldots 500$,
which is well reproduced by the semiclassical prediction based on a $9$:$1$
resonance inside the islands.

In conclusion, we have presented a straightforward semiclassical scheme to
reproduce tunneling rates between regular islands in mixed systems.
Our approach is based on the existence of a prominent nonlinear resonance
inside the island, and uses elementary classical parameters associated with
this resonance to estimate the coupling rate from the island to the chaos.
In combination with a random matrix description of the chaotic part of the
Hamiltonian, we obtain a simple expression for the average level splittings
between symmetry-related islands, which agrees reasonably well with the exact
quantum splittings calculated for the kicked rotor and the kicked Harper.
Our study underlines that nonlinear resonances govern the coupling between
regular islands and the surrounding chaotic sea in the semiclassical limit.
We expect that they play an equally prominent role also in complex systems
with more degrees of freedom, for which our approach could develop into a
useful method to quantitatively estimate tunneling rates in presence of chaos.

We thank O.~Brodier, S.~Keshavamurthy, S.~Tomsovic, and D.~Ullmo for fruitful
and inspiring discussions.
Support from the Deutsche Forschungsgemeinschaft is gratefully acknowledged.


\begin{thebibliography}{10}

\bibitem{DavHel81JCP}
M.~J. Davis and E.~J. Heller, J. Chem. Phys. {\bf 75},  246  (1981).

\bibitem{Cre98}
S. Creagh,  in {\em Tunneling in Complex Systems}, edited by S. Tomsovic (World
  Scientific, Singapore, 1998), p.\ 1.

\bibitem{LinBal90PRL}
W.~A. Lin and L.~E. Ballentine, Phys. Rev. Lett. {\bf 65},  2927  (1990);
F. Grossmann, T. Dittrich, P. Jung and P. H{\"a}nggi, Phys. Rev. Lett. {\bf
  67},  516  (1991).

\bibitem{BohTomUll93PR}
O. Bohigas, S. Tomsovic, and D. Ullmo, Phys. Rep. {\bf 223},  45  (1993);
O. Bohigas, D. Boos\'{e}, R. {Egydio de Carvalho}, and V. Marvulle, Nucl. Phys.
  A {\bf 560},  197  (1993).

\bibitem{TomUll94PRE}
S. Tomsovic and D. Ullmo, Phys. Rev. E {\bf 50},  145  (1994).

\bibitem{DorFri95PRL}
E. Doron and S.~D. Frischat, Phys. Rev. Lett. {\bf 75},  3661  (1995);
S.~D. Frischat and E. Doron, Phys. Rev. E {\bf 57},  1421  (1998).

\bibitem{ZakDelBuc98PRE}
J. Zakrzewski, D. Delande, and A. Buchleitner, Phys. Rev. E {\bf 57},  1458
  (1998).

\bibitem{NoeSto97N}
J.~U. N{\"o}ckel and A.~D. Stone, Nature {\bf 385},  45  (1997).

\bibitem{DemO00PRL}
C. Dembowski {\it et~al.}, Phys. Rev. Lett. {\bf 84},  867  (2000).

\bibitem{MouO01PRE}
A. Mouchet {\it et~al.}, Phys. Rev. E {\bf 64},  016221  (2001).

\bibitem{HenO01N}
W.~K. Hensinger {\it et~al.}, Nature {\bf 412},  52  (2001);
D.~A. Steck, W.~H. Oskay, and M.~G. Raizen, Science {\bf 293},  274  (2001).

\bibitem{LeyUll96JPA}
F. Leyvraz and D. Ullmo, J. Phys. A {\bf 29},  2529  (1996).

\bibitem{ShuIke95PRL}
A. Shudo and K.~S. Ikeda, Phys. Rev. Lett. {\bf 74},  682  (1995);
Phys. Rev. Lett. {\bf 76},  4151  (1996);
T. Onishi, A. Shudo, K.~S. Ikeda, and K. Takahashi, Phys. Rev. E {\bf 64},
  025201  (2001).

\bibitem{PodNar03PRL}
V.~A. Podolskiy and E.~E. Narimanov, Phys. Rev. Lett. {\bf 91},  263601
  (2003).

\bibitem{Ozo84JPC}
A.~M. {Ozorio de Almeida}, J. Phys. Chem. {\bf 88},  6139  (1984).

\bibitem{BonO98PRE}
L. Bonci, A. Farusi, P. Grigolini, and R. Roncaglia, Phys. Rev. E {\bf 58},
  5689  (1998).

\bibitem{BroSchUll01PRL}
O. Brodier, P. Schlagheck, and D. Ullmo, Phys. Rev. Lett. {\bf 87},  064101
  (2001); Ann. Phys. {\bf 300},  88  (2002).

\bibitem{Kes03JCP}
S. Keshavamurthy, J. Chem. Phys. {\bf 119},  161  (2003).

\bibitem{LicLie}
A.~J. Lichtenberg and M.~A. Lieberman, {\em Regular and Stochastic Motion}
  (Springer-Verlag, New York, 1983).

\bibitem{LebO90PRL}
P. Leboeuf, J. Kurchan, M. Feingold, and D.~P. Arovas, Phys. Rev. Lett. {\bf
  65},  3076  (1990).

\bibitem{GeiRadRub86PRL}
T. Geisel, G. Radons, and J. Rubner, Phys. Rev. Lett. {\bf 57},  2883  (1986);
N.~T. Maitra and E.~J. Heller, Phys. Rev. E {\bf 61},  3620  (2000).

\bibitem{KetO00PRL}
R. Ketzmerick, L. Hufnagel, F. Steinbach, and M. Weiss, Phys. Rev. Lett. {\bf
  85},  1214  (2000).

\end{thebibliography}
\end{document}